\documentclass[aps,prd,twocolumn,preprintnumbers,superscriptaddress,dblfloatfix,nofootinbib]{revtex4-1}
\usepackage{bbm}
\usepackage{mathrsfs}
\usepackage{slashed}
\usepackage{caption}
\usepackage{epstopdf}
\usepackage[normalem]{ulem}
\usepackage[bottom]{footmisc}
\usepackage{subcaption}
\usepackage{bbold}
\usepackage{threeparttable}
\usepackage{booktabs}
\usepackage{changepage}
\usepackage[utf8]{inputenc}

\usepackage{grffile}

\usepackage{graphicx}  % needed for figures
\usepackage{dcolumn}   % needed for some tables
\usepackage{bm}        % for math
\usepackage{amssymb}   % for math
\usepackage{setspace}
\usepackage{amsmath, amssymb, setspace}
\usepackage{array}
\usepackage{booktabs}
\usepackage{caption}
\usepackage{indentfirst}
\usepackage{float}
\usepackage{lmodern}
\usepackage{multirow}
\usepackage{soul}
\usepackage[normalem]{ulem}

\usepackage{braket}
\usepackage{comment}

\usepackage{adjustbox} 
\newcommand{\ie}{\emph{i.e.~}}

\usepackage[dvipsnames]{xcolor}
\usepackage[normalem]{ulem}

\newcommand{\Fig}[1]{Fig.~\ref{#1}}

\newcommand{\Tab}[1]{Table~\ref{#1}}

\usepackage{color}
\begin{document}

%\preprint{\%}

\title{Pre-Supernova Neutrinos in Large Dark Matter Direct Detection Experiments}

\author{Nirmal Raj}
\email[]{nraj@triumf.ca}
\affiliation{TRIUMF, 4004 Wesbrook Mall, Vancouver, BC V6T 2A3, Canada}

\author{Volodymyr Takhistov}
\email[]{vtakhist@physics.ucla.edu}
\affiliation{Department of Physics and Astronomy, University of California, Los Angeles\\
Los Angeles, CA 90095-1547, USA}

\author{Samuel J. Witte}
\email[]{sam.witte@ific.uv.es}
\affiliation{Instituto de F\'{i}sica Corpuscular (CSIC-Universitat de Val\`{e}ncia), Paterna (Valencia), Spain}
 
%%%%%%%%%%%%%%%%%%%%%%%%%%%

\date{\today}

\begin{abstract}
The next Galactic core-collapse supernova (SN) is a highly anticipated observational target for neutrino telescopes. 
However, even prior to collapse, massive dying stars shine copiously in ``pre-supernova'' (pre-SN) neutrinos, which can potentially act as efficient SN warning alarms and provide novel information about the very last stages of stellar evolution. We explore the sensitivity to pre-SN neutrinos of large scale direct dark matter detection experiments, which, unlike dedicated neutrino telescopes, take full advantage of coherent neutrino-nucleus scattering.
We find that argon-based detectors with target masses of $\mathcal{O}(100)$ tonnes (i.e.~comparable in size to the proposed ARGO experiment) operating at sub-keV thresholds can detect $\mathcal{O}(10-100)$ pre-SN neutrinos coming from a source at a characteristic distance of $\sim$200 pc, such as Betelgeuse ($\alpha$ Orionis). 
Large-scale xenon-based experiments with similarly low thresholds could also be sensitive to pre-SN neutrinos.
For a Betelgeuse-type source, large scale dark matter experiments could provide a SN warning siren $\sim$10 hours prior to the explosion. We also comment on the complementarity of large scale direct dark matter detection experiments and neutrino telescopes in the understanding of core-collapse SN. 
\end{abstract}

\maketitle
%%%%%%%
\section{Introduction}
%%%%%%%%

Stars with mass $\gtrsim 8 M_{\odot}$ and that ignite nuclear fuel burning non-explosively, explode as core-collapse supernovae (SNe) at the end of their lifetime (see Ref.~\cite{Burrows:2012ew} for a review), leaving behind a compact remnant. 
As the star's nuclear fuel becomes exhausted, the inner core collapses under gravity. 
Contraction of the core leads to a dramatic increase in density, with nuclear forces halting and bouncing back the rapid collapse, resulting in a propagating outward shockwave. 
Copious emission of $\sim$10$-$30 MeV neutrinos within a $\sim$10 second burst are a generic byproduct of this event, typically carrying away $\sim10^{53}$ erg of the original star's gravitational binding energy. The luminosity in neutrinos from core-collapse supernovae  greatly exceeds its optical counterpart. The general picture of the outlined supernova mechanism was decisively confirmed by the observation of neutrinos from SN 1987A~\cite{Bionta:1987qt,Hirata:1987hu,Alekseev:1988gp}.
With the expected rate of Galactic core-collapse SNe of around few per century, anticipation and preparation for future SN observations is a principal goal of large-scale neutrino experiments~\cite{Mirizzi:2015eza}, however large scale direct dark matter detection experiments will also be sensitive to SNe neutrinos in a highly complementary manner~\cite{Chakraborty:2013zua,XMASS:2016cmy,Lang:2016zhv,Kozynets:2018dfo,Khaitan:2018wnf}.

 Even prior to inception of the core-collapse, a significant emission of  $\sim$ MeV ``pre-supernova'' (pre-SN) neutrinos is expected from the supernova progenitor during the final nuclear fuel burning stages,  in particular from the silicon (Si) burning~\cite{Odrzywolek:2003vn,Kato:2015faa,Yoshida:2016imf,Patton:2017neq}.
The resulting pre-SN neutrino luminosity is typically smaller than that of SN neutrinos by a few orders of magnitude.
Detection of pre-SN neutrinos will directly probe the very late stages of nuclear fusion processes beyond hydrogen and helium within the SN system, providing vital information about the temperature and density near the star's core at that time.
Furthermore, these neutrinos could provide an early supernova warning trigger, dramatically improving upon the current Supernova Early Warning System  (SNEWS) \cite{Antonioli:2004zb} network.

Sensitivity studies for pre-SN neutrinos have been conducted~\cite{Odrzywolek:2003vn,Kato:2015faa,Yoshida:2016imf,Patton:2017neq} for current and future neutrino experiments, primarily focusing on liquid scintillator-based (Borexino~\cite{Alimonti:2008gc},
 KamLAND~\cite{Gando:2013nba,Gando:2014wjd},  SNO+~\cite{Andringa:2015tza},
 JUNO~\cite{An:2015jdp, Han:2015roa}) and water Cherenkov-based  (SNO~\cite{Aharmim:2011vm}, Super-Kamiokande~\cite{Fukuda:2002uc,Abe:2013gga} and Hyper-Kamiokande~\cite{Abe:2018uyc}, including gadolinium dissolution~\cite{Beacom:2003nk}) detectors. 
 A dedicated study by the KamLAND collaboration has been also recently carried out~\cite{Asakura:2015bga}. 
 These analyses, however, primarily focused on detection via the inverse beta decay (IBD), $\overline{\nu}_e + p \rightarrow e^+ + n$ channel, which has a kinematic threshold of neutrino energy  = 1.8 MeV and is limited to $\overline{\nu}_e$ interactions.

Among the major open questions in modern physics is the nature of dark matter (DM). 
For many decades a leading DM candidate has been the weakly interacting massive particle (WIMP), however despite a dedicated and multi-pronged experimental search program, WIMPs have remained elusive. 
Some of the most stringent constraints on WIMP DM come from direct detection experiments, which search primarily for rare neutral current interactions with nuclei in deep underground laboratories.
Future generations of these experiments will continue to carve out WIMP parameter space, eventually encountering an irreducible background arising from coherent scattering of neutrinos produced from: the Sun, atmospheric interactions, SNe, the interior of the Earth, etc. --- collectively, these interactions constitute what is called the ``neutrino floor''~\cite{Billard:2013qya,Monroe:2007xp,Gelmini:2018ogy}.
Since neutrinos can mimic the DM signal, they limit the DM discovery sensitivity.
However, the neutrino signal, which definitely exists and will be observed, constitutes in itself an interesting subject of study. 
It is thus vital to further explore detection capabilities of direct detection experiments beyond particle DM.

Large direct detection experiments are themselves effective neutrino detectors, capable of probing neutrinos in regimes complementary to those studied with conventional neutrino experiments. 
One reason for this is the very low detection thresholds achievable in direct detection experiments, often around or below keV level. 
Furthermore, with heavy nuclei as detector targets, these experiments achieve high detection rates via coherent neutrino-nucleus scattering, whose cross-section scales approximately as neutron number squared. This process has been recently directly observed~\cite{Akimov:2017ade}. 
Both coherent neutrino-nucleus scattering and elastic neutrino-electron scattering have been considered in a range of studies related to neutrino physics for direct detection experiments, including
sterile neutrinos (e.g.~\cite{Pospelov:2011ha,Billard:2014yka}),
non-standard neutrino interactions (e.g.~\cite{Harnik:2012ni,Dutta:2017nht}),
solar neutrinos (e.g.~\cite{Billard:2014yka,Schumann:2015cpa,Franco:2015pha,Cerdeno:2017xxl,Essig:2018tss,Newstead:2018muu}),  
geoneutrinos~\cite{Gelmini:2018gqa}, 
neutrinos from DM annihilations and decays~\cite{Cherry:2015oca,Cui:2017ytb,McKeen:2018pbb}, as well as 
supernova neutrinos~\cite{Chakraborty:2013zua,XMASS:2016cmy,Lang:2016zhv,Kozynets:2018dfo,Khaitan:2018wnf}.

In this work we explore pre-SN neutrino detection capabilities of large direct detection experiments. 
Unlike traditional neutrino detectors using IBD, coherent-neutrino interactions allow for direct detection experiments to be unconstrained by the IBD kinematic threshold on neutrino energy, and also to have sensitivity to all six ($\nu_e,\overline{\nu}_e,\nu_\mu,\overline{\nu}_\mu,\nu_\tau,\overline{\nu}_\tau$) neutrino flavors. As pre-SN neutrinos are significantly softer than SN neutrinos and also fewer in number, efficient detection entails a low energy threshold and a large target mass.  As we will show, a favorable combination of the above is achievable in a future argon-based $\mathcal{O}(100)$ ton-scale detector, such as the recently proposed ARGO experiment~\cite{zuzel2017darkside,Aalseth:2017fik}. Large-scale dark matter direct detection experiments are therefore complementary to neutrino detectors and are insensitive to uncertainties associated with neutrino oscillations.
 
This paper is organized as follows. 
 In Sec.~\ref{sec:dd} we review the capabilities of future dark matter direct detection experiments, with emphasis on their ability to achieve low energy thresholds suitable for pre-SN neutrino detection.
 In Sec.~\ref{sec:sigbg} we show the differential fluxes of pre-SN neutrinos and background neutrinos, and obtain their event rates at direct detection experiments. In Sec.~\ref{sec:nucrecoil} we describe the expected nuclear recoil rates.
 In Sec.~\ref{sec:sensitivity} we derive our sensitivities to pre-SN neutrinos, showing that $\mathcal{O}(100)$ ton-scale argon detector (e.g.~ARGO~\cite{zuzel2017darkside,Aalseth:2017fik}) can constitute an efficient target.
 In Sec.~\ref{sec:summary} we summarize and conclude.

%%%%%%%%%%%%%% 
\section{Large Direct Detection Experiments}
\label{sec:dd}
%%%%%%%%%%%%%%

A variety of proposals for the next generation of direct detection experiments has been suggested~\cite{Cushman:2013zza}.
As it is difficult to predict the exact final design and hence the associated detection capabilities, we consider optimistic but realistic detector configurations based on current technology.
The detector target elements we consider are argon (Ar), xenon (Xe), germanium (Ge)  and silicon (Si), which we list in~\Tab{tab:targetmat} with their respective isotope abundance\footnote{We do not include isotopic abundances of elements that contribute less than $1\%$. Note that we refer to naturally occurring abundances, while isotopically-modified target material could be used. We do not expect this to significantly alter our conclusions.}.

\begin{table}[tb]
	\begin{center}
		\vspace*{0cm}
		\begin{tabular}[c]{l|cc} \hline\hline
			Target   & $A(Z)$ & Isotope  \\ 
			Material & & Fraction  \\
\hline
		    xenon(Xe)  & 128(54)  & 0.019   \\
		               & 129(54)  & 0.264       \\
		               & 130(54)  & 0.041       \\
		               & 131(54)  & 0.212       \\
		               & 132(54)  & 0.269       \\
		               & 134(54)  & 0.104      \\
		               & 136(54)  & 0.089       \\
\hline
		 argon(Ar)     & 40(18)  & 0.996    \\
\hline
		germanium(Ge)  & 70(32)  & 0.208    \\
		               & 72(32)  & 0.275     \\
		               & 73(32)  & 0.077      \\
		               & 74(32)  & 0.363      \\
		               & 76(32)  & 0.076      \\
\hline
		  silicon(Si)  & 28(14)  & 0.922   \\
		               & 29(14)  & 0.047    \\
		               & 30(14)  & 0.031     \\
		          
\hline\hline
		\end{tabular}
		\caption{Summary of target materials considered for our experimental configurations.}
		\label{tab:targetmat}
		\vspace*{-0.5cm}
	\end{center}
\end{table}

Throughout this work, we optimistically assume that detectors can achieve perfect detection threshold and energy resolution.
Furthermore, we assume that our background originates either from neutrino interactions alone, as might optimistically be the case should conventional backgrounds be reduced to a point where they can be neglected, or from electronic recoils in the detector, which allows us to to explore the possibility that  reducible backgrounds dominate over those coming from solar neutrinos. The former assumption allows one to treat different experimental configurations on the same footing, independent of specific detector design choices, while the latter allows us to assess the extent to which electronic backgrounds need to be reduced in order to extract useful information from the  pre-SN signal. As significant non-neutrino backgrounds are expected, it would be difficult to justify fully neglecting electronic backgrounds for neutrino-electron scattering signals. 
For these reasons, and because the neutrino-electron scattering rate is orders of magnitude smaller than the coherent neutrino-nucleus scattering rate (see, e.g. Ref.~\cite{Lang:2016zhv}),
we will focus solely on neutrino-nucleus scattering in this work.

We assume our detectors to be located at SNOLAB (Sudbury, Canada), which will likely host a number of next-generation direct detection experiments. However, we emphasize that this assumption will not significantly influence any of our conclusions as the strong time-dependence of the pre-SN neutrino flux allows for unambiguous separation of signal events from  background.
The depth of this lab (6010 m.w.e.) ensures that backgrounds due to cosmogenic muons are highly suppressed.

%%%%%%%%%%%%%%%%
\begin{figure}[tb]
\centering
\includegraphics[width=.45\textwidth]{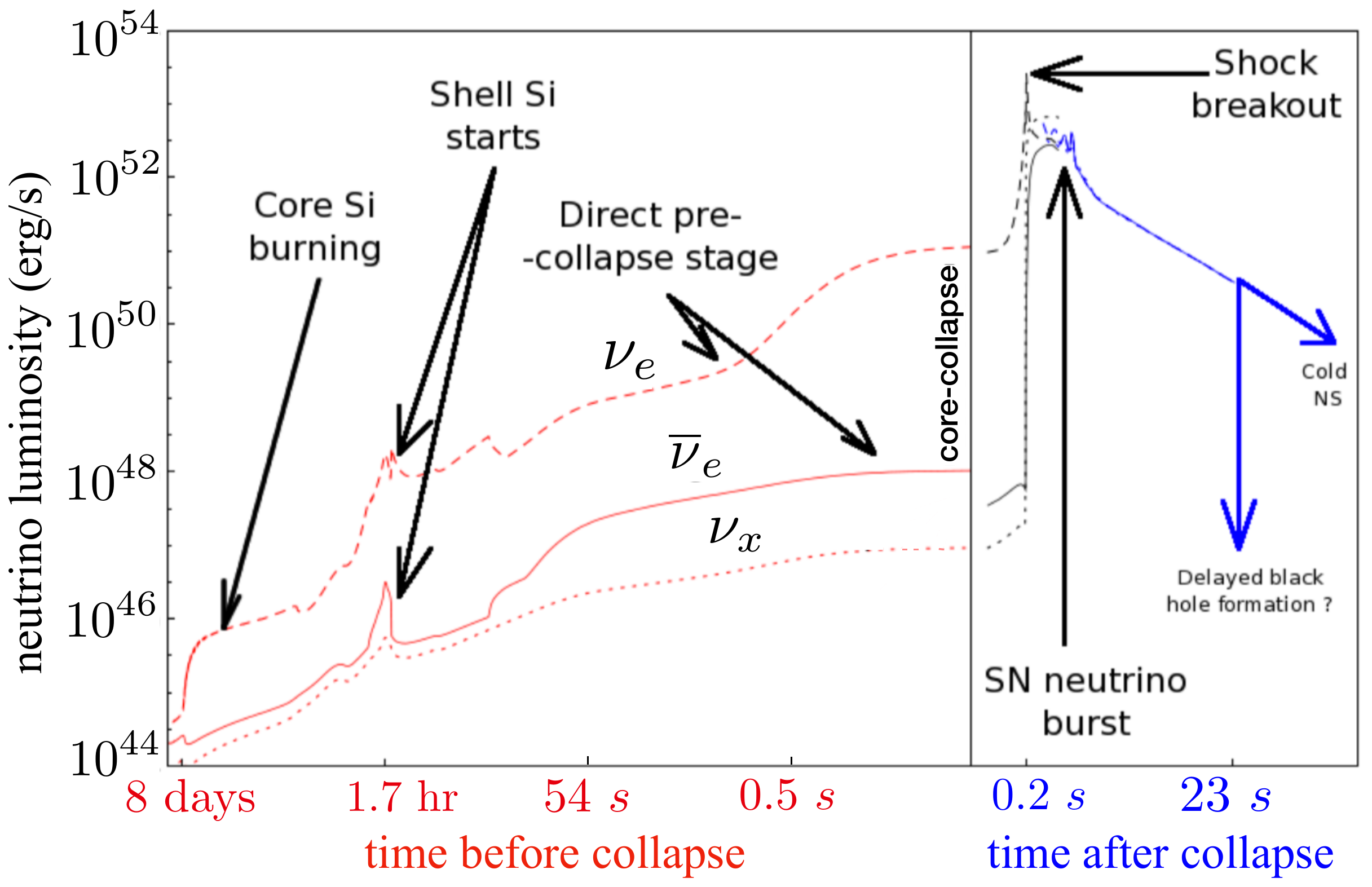}
\caption{\label{fig:presnsignal} 
Evolution of the neutrino luminosity of a 15~$M_{\odot}$ star in the final stages of its life.
Contributions from electron neutrinos ($\nu_e$), electron anti-neutrinos ($\overline{\nu}_e$) and the averaged contribution from $\nu_\mu$, $\overline{\nu}_\mu$, $\nu_\tau$, and $\overline{\nu}_\tau$ flavors (denoted by ``$\nu_x$") are shown, for both before supernova core-collapse (pre-SN neutrinos) and after core-collapse (SN neutrinos).
Note the change of scale on the x-axis after core-collapse.
This figure is adapted from Ref.~\cite{Odrzywolek:2010zz}.
}
\end{figure}
%%%%%%%%%%%%%%%

\begin{figure}[tb]
\includegraphics[width=.45\textwidth]{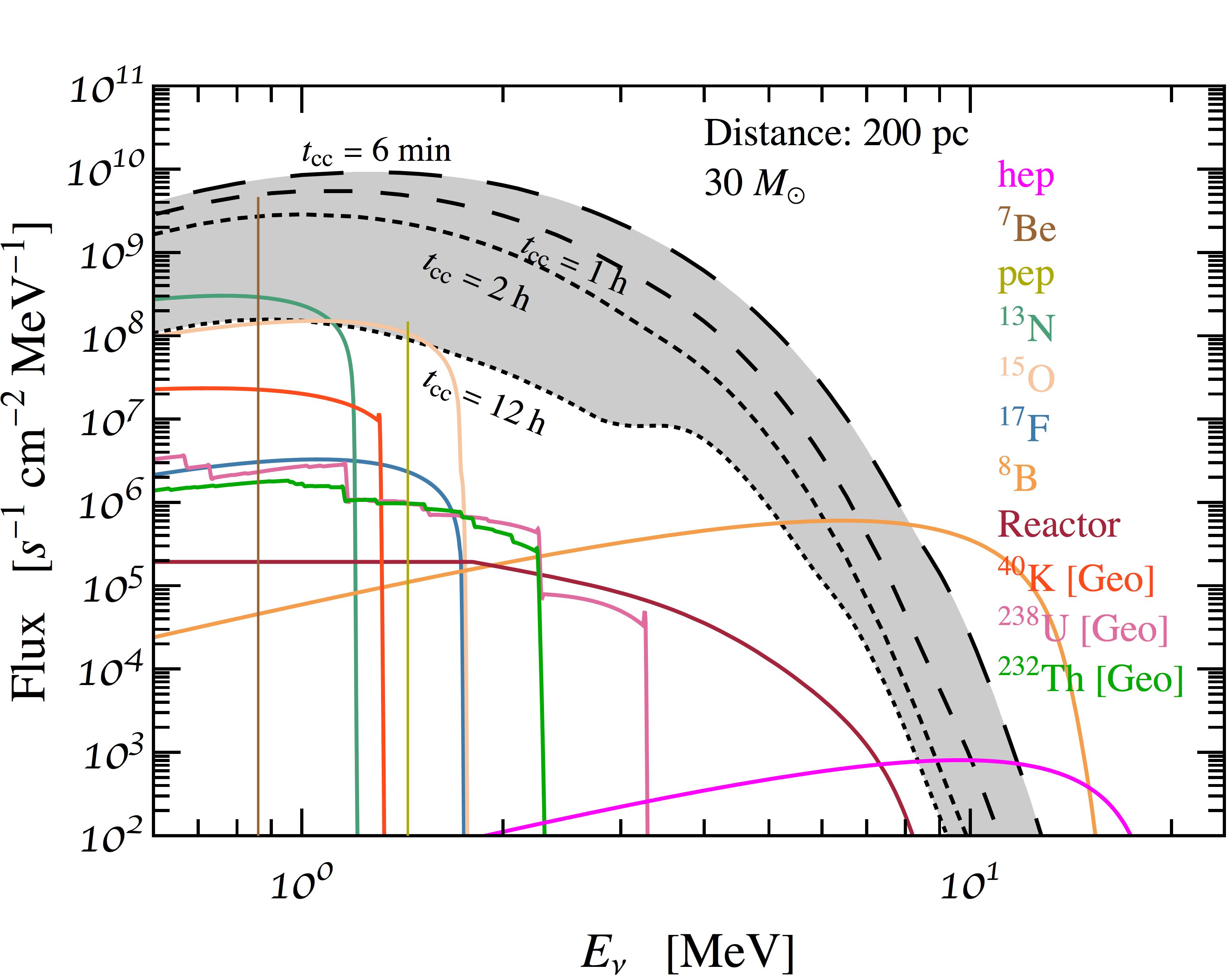}
\caption{\label{fig:neutrino_flux}
 Neutrino fluxes at the SNOLAB location.
The pre-supernova neutrino signal flux (dashed curves) is evaluated at various times $t_{cc}$ prior to
core-collapse. 
The backgrounds (solid curves) originate from solar, reactor and geo-neutrinos. 
Their fluxes and maximum energies are provided in Table~\ref{tab:backflux}. 
The solar $^8$B neutrinos constitute our main background.}   
\end{figure}
%%%%%%%%%%%%%%%%%%

For future large-scale direct detection experiments, we adopt fiducial target masses characteristic of proposed generation-3 experiments and thresholds consistent with those achievable in current and near-future experiments.
In particular,
we assume that the argon-based experiment will have detector mass of 300 tonnes, as proposed for ARGO~\cite{zuzel2017darkside,Aalseth:2017fik}. 
For xenon, we take a detector mass of 50 tonnes, as proposed for DARWIN~\cite{Aalbers:2016jon}.
For silicon and germanium we assume target mass of 50 kg, similar to what is proposed for SuperCDMS-SNOLAB~\cite{Agnese:2016cpb}.

We briefly elaborate on the energy thresholds assumed in our analysis.
In a recent search for light DM, the argon-based DarkSide-50 experiment has demonstrated the ability to reach a nuclear recoil threshold as low as 0.6 keV while maintaining low backgrounds, although electronic backgrounds in this low threshold analysis cannot be fully mitigated\footnote{A broader search for DM at DarkSide-50 employs a somewhat higher threshold~\cite{Agnes:2018fwg}.}~\cite{Agnes:2018ves}.
For xenon-based detectors, such as
XENON1T~\cite{Aprile:2017iyp,Aprile:2018dbl},
LUX~\cite{Akerib:2016vxi} and 
PandaX-II~\cite{Cui:2017nnn}, typical background-free searches operate efficiently at recoil energies above $\sim 5$ keV. However, a nuclear recoil threshold of 0.7 keV has been recently achieved~\cite{Aprile:2016wwo} at the expense of a reduced efficiency and higher background.
For both argon and xenon, these sub-keV thresholds were accomplished using information from only the $S$2 signal, produced when electrons are drifted out of the liquid and into the gas phase of the time projection chamber (TPC). 
The fallback of adopting a $S$2-only analysis is the loss of pulse shape discrimination that is necessary for a strong rejection of electronic recoils. 
Without any additional experimental techniques, such backgrounds are currently at a level that would plague pre-SN detection. 
However, there are significant promising experimental R\&D efforts  to reduce and better understand the characteristics of these background sources that are currently underway, suggesting that experimental setups with dramatically reduced backgrounds could be feasible by the time generation-3 experiments are operational. Nevertheless, we will present results in Sec.~\ref{sec:sensitivity} assuming electronic recoil backgrounds based on measurements from available experiments. We note that the $S$2-only analyses performed in these experiments typically operate at reduced efficiencies. For example, by comparing the standard analyses performed by DarkSide-50 and Xenon-100 with the $S$2-only counterpart, the associated efficiencies are only $\sim 40\%$ and $15\%$, respectively. In order to compare these experiments in a more balanced manner, we neglect this efficiency reduction and simply note that this may affect the observable number of neutrinos in argon and xenon by a factor of $\sim 2$ and $\sim 7$, respectively.

Both germanium and silicon will be used in the future SuperCDMS-SNOLAB experiment. 
The SuperCDMS collaboration has recently determined that recoil thresholds as low as 78 eV are achievable with their high voltage (HV) silicon detectors, however at the moment their estimation of a 40 eV threshold in the germanium HV detector relies on the use of Lindhard theory  and has yet to be directly verified (see Table VII of Ref.~\cite{Agnese:2016cpb}). We  emphasize that the HV detectors are subject to larger backgrounds than the conventional detectors, with thresholds closer to the level of $\mathcal{O}(200)$ eV~\cite{Agnese:2016cpb}.

In Table~\ref{tab:experiments} we summarize the thresholds and target masses for our experimental configurations.
Note that in this table and throughout this work our threshold values denote the energy deposited, so that keV denotes keV$_{\rm r}$, as opposed to the electron equivalent energy (keV$_{\rm ee}$).
The sensitivities to pre-SN neutrinos of near-future generation-2 experiments, such as the argon-based DarkSide-20k~\cite{Aalseth:2017fik} and the xenon-based XENONNT~\cite{Aprile:2015uzo} and LZ~\cite{Akerib:2018lyp}, can be simply inferred by rescaling our results by the corresponding target masses as appropriate.

\section{Pre-Supernova Neutrino Signal and Backgrounds}
\label{sec:sigbg}

\subsection{Pre-supernova neutrino flux}

When the fuel of a particular element in the core of a star is exhausted, outward thermal pressure forces can no longer counterbalance those induced by gravity. 
This causes the star to contract. 
The compression-induced increase in density and temperature can result in fusion of heavier elements. 
After nuclear burning within the core, the same element can subsequently burn in the outer shell layers. 
The later burning stages occur on increasingly shorter time scales. Energy loss can occur from photo-emission from the outer surfaces as well as neutrino-emission  from within the star. After the helium-burning stage, energy losses are dominated by neutrinos produced within the interior. For stars sufficiently massive to reach the silicon burning stage, the temperatures and densities are so high that electron capture is rampant, resulting in a high flux of $\nu_e$ and overall neutronization. This continues until the onset of the core-collapse and subsequent shock break-out, which is accompanied by copious neutrino emission. 

A variety of thermal and $\beta$ processes contribute to the overall pre-SN neutrino emission, including
pair-annihilation ($e^+ + e^- \rightarrow \nu + \overline{\nu}$),
plasmon decays ($\gamma^{\ast} \rightarrow \nu +\overline{\nu}$), 
photo-neutrino production ($e^{\pm} + \gamma \rightarrow e^{\pm} + \nu + \overline{\nu}$), $e^{\pm}$ capture and
and $\beta^{\pm}$ decays. 
Non-standard neutrino effects could also affect
stellar cooling~\cite{Heger:2008er}.

The evolution of the pre-SN neutrino flux (compared with the SN neutrino flux) is displayed in Fig.~\ref{fig:presnsignal}, adapted from Ref.~\cite{Odrzywolek:2010zz}. 
While the prediction of the pre-SN neutrino flux is not without modeling uncertainties, independent numerical simulations have found closely matching results.
These uncertainties are subdominant compared to the uncertainties on the measurements of the distances to source stars.
For our analysis we use the tabulated neutrino fluxes for $M = 15M_{\odot}$ and $M = 30M_{\odot}$ stars as computed by Patton~\textit{et al.}~\cite{Patton:2017neq}. 
We have confirmed that using the fluxes computed by Odrzywolek~\textit{et al.}~\cite{Odrzywolek:2003vn} does not significantly alter our results.

\subsection{Neutrino backgrounds}
\label{ss:nuback}

Since pre-SN neutrinos carry $\sim$ MeV energies, the relevant backgrounds are due to solar, reactor and geo-neutrinos (see Fig.~\ref{fig:neutrino_flux}), which dominate at energies $E_{\nu} \lesssim 10$ MeV. 
Among these, reactor and geo-neutrinos fluxes depend sensitively on the detector location. 
We have summarized these backgrounds in Table~\ref{tab:backflux}.
Although our main background originates from solar $^8$B neutrinos, which has significant flux even for neutrino energies beyond 10 MeV, we also comment on the other background contributions for completeness.

%%%%%%%%%%%%%%%%
\begin{table}[t]
	\begin{center}  \setlength{\extrarowheight}{2pt}
		\vspace*{0cm}
		\begin{tabular}[c]{l|cc} \hline\hline
			Neutrino   & Flux & Max Energy\\ 
			Component & $\mathrm{[cm^{-2} s^{-1}]}$ & $\mathrm{E_{\nu}~[MeV]}$
			\\\hline
			$\mathrm{Solar~(\nu_e, pp)}$  & ~$6.03(1\pm0.006) \times 10^{10}$  & $0.42$     \\
			$\mathrm{Solar~(\nu_e, pep [line])}$  & $1.47(1\pm0.012) \times 10^{8}$  & $1.45$     \\
			$\mathrm{Solar~(\nu_e, hep)}$  & $8.31(1\pm0.300) \times 10^{3}$  & $18.77$     \\
			$\mathrm{Solar~(\nu_e, {}^{7}Be [line~1])}$  & $4.56(1\pm0.070) \times 10^{8}$  & $0.39$     \\
			$\mathrm{Solar~(\nu_e, {}^{7}Be [line~2])}$  & $4.10(1\pm0.070) \times 10^{9}$  & $0.87$     \\
			Solar~($\nu_e$, $^8$B)  & $4.59(1\pm0.140) \times 10^{6}$  & $16.80$     \\
		    $\mathrm{Solar~(\nu_e, ^{13}N)}$  & $2.17(1\pm0.140) \times 10^{8}$  & $1.20$ \\
		    $\mathrm{Solar~(\nu_e, {}^{15}O)}$  & $1.56(1\pm0.150) \times 10^{8}$  & $1.73$ \\
		    $\mathrm{Solar~(\nu_e, {}^{17}F)}$  & $3.40(1\pm0.170) \times 10^{6}$  & $1.74$ \\ 
\hline
		    $\mathrm{Reactor~(\overline{\nu}_e})$  & $ 5.96 (1\pm0.080) \times 10^{5}$  & $10.00$ \\
		    \hline
		    Geo~($\overline{\nu}_e$, $^{40}$K)  & $2.19 (1 \pm  0.168) \times 10^{7}$  &   1.32 \\
Geo~($\overline{\nu}_e$, $^{238}$U)  &  $4.90 (1 \pm  0.200) \times 10^{6}$ &   ~~3.99\footnote{In \Fig{fig:neutrino_flux}, the $^{238}$U flux appears to cutoff at much lower energies, however this is simply a consequence of not extending the y-axis to lower values. See Fig. 1 of~Ref.~\cite{Gelmini:2018ogy} for an extended spectrum. }\\
Geo~($\overline{\nu}_e$, $^{232}$Th)  & $4.55 (1 \pm  0.257) \times 10^{6}$  &   2.26 \\
\hline\hline
		\end{tabular}
		\end{center}
		\caption{Fluxes and maximum energies of the neutrino backgrounds to the pre-supernova neutrino signal in direct detection experiments. 
		These consist of solar, reactor and geo-neutrino contributions. 
		The solar $^8$B neutrinos constitute our main background.}
		\label{tab:backflux}
		\vspace*{-0.5cm}
\end{table}
%%%%%%%%%%%%%%%%

\subsubsection{Solar neutrinos}

Electron neutrinos that are produced via nuclear fusion reactions in the Sun contribute predominantly to the neutrino background at low energies (see Ref.~\cite{Robertson:2012ib} for a review).
 This background is independent of the laboratory location.
The solar neutrino flux is composed of contributions from multiple reaction chains, with varying resultant fluxes and energies.
The proton-proton cycle contributes to more than 98\% of the energy flux. 
Initiated by the reaction $p + p \rightarrow ~^{2}{\rm H} + e^+ + \nu_e$, this cycle gives rise to $pp$, $hep$, $pep$, $^{7}$Be and $^{8}$B neutrinos.
The carbon-nitrogen-oxygen (CNO) cycle accounts for the rest of the Sun's energy, giving rise to $^{13} $N, $^{15}$O and $^{17}$F neutrinos.

Predictions of solar neutrino fluxes depend on the solar model.
The Standard Solar Model (GS98)~\cite{Grevesse:1998bj} has been demonstrated to agree well with helioseismological studies. 
However, more modern models such as AGSS09~\cite{Asplund:2009fu} that are more internally consistent, show lesser degree of agreement with the helioseismology results, leading to a discrepancy known as the ``solar metallicity'' problem. 
In this work we will assume the solar neutrino fluxes and uncertainties as predicted by AGSS09 (see Table 2 of \cite{Robertson:2012ib}).
These are provided in \Tab{tab:backflux}. 
Our dominant background originates from $^8$B neutrinos. Since the $^8$B flux difference between the two solar models is only $\sim 20$\%, using a different solar model will not significantly affect our results. Future measurements of the solar neutrino flux will further reduce this uncertainty.

\subsubsection{Reactor neutrinos}

Fission $\beta$-decays of uranium ($^{235}$U and $^{238}$U) and plutonium ($^{239}$Pu and $^{241}$Pu) in reactor fuels give rise to reactor electron anti-neutrinos (see Ref.~\cite{Hayes:2016qnu} for a review).
The corresponding neutrino flux depends sensitively on the reactor operation since these isotopes are short-lived, and since the reactor's fuel composition and relative isotope fractions evolve with time. The reactor neutrino background depends on the location of the direct detection experiment, i.e. on the specifics of and distances to nearby reactors. 
For the SNOLAB location, we calculate this background using the formalism and reactors described in Ref.~\cite{Gelmini:2018ogy}.

\subsubsection{Geo-neutrinos}

Geo-neutrinos are predominantly electron anti-neutrinos\footnote{We note that electron neutrinos are also produced in subdominant quantities (e.g. from electron capture in $^{40}$K, contributing at the level of $\sim 11\%$). We neglect this contribution here as geoneutrinos are not expected to noticeably impede the detection of pre-SN neutrinos.} originating from the $\beta$-decay branches of the Earth's major heat-producing nuclear reactions, involving isotopes of 
potassium ($^{40}$K), 
thorium ($^{232}$Th),
and uranium ($^{238}$U).
Recently, KamLAND~\cite{Araki:2005qa} and Borexino~\cite{Bellini:2010hy} have observed a geoneutrino flux.
We take the spectrum for each of these elements from Ref.~\cite{Araki:2005qa}. 
The respective location-dependent total flux is predicted from a geophysics-based three-dimensional global Earth model of heat-producing element distribution \cite{huang:2013geomodel}. 

\section{Nuclear Recoil rates}
\label{sec:nucrecoil}
 
 The Standard Model neutrino-nucleus coherent scattering cross section is given by~\cite{Freedman:1977xn} 
%%%%%%%%%%%%%%%%
\begin{equation}\label{eq:nucxsec}
\dfrac{d \sigma^I}{d E_r} (E_{\nu}, E_r) = \dfrac{G_F^2 m_I}{4 \pi} Q_W^2 \left(1 - \dfrac{m_I E_r}{2 E_{\nu}^2}\right) F_{I}^2 (E_r)~,
\end{equation}
%%%%%%%%%%%%%%%%
where $m_I$ is the target nuclide mass, $G_F = 1.1664~\times~10^{-5}$~GeV$^{-2}$ is the Fermi coupling constant, $F_I (E_r)$ is the form factor, taken to be the Helm form factor~\cite{Helm:1956zz}, $Q_W = [(1 - 4 \sin^2 \theta_{\rm W}) Z_I-N_I]$ is the weak nuclear charge, $N_I$ is the number of neutrons, $Z_I$ is the number of protons, and $\theta_{\rm W}$ is the Weinberg angle. Since $\sin^2 \theta_{\rm W} = 0.23867$ at low energies \cite{Erler:2004in}, the coefficient in front of $Z_I$ in $Q_W$ approximately vanishes, and the coherent neutrino-nucleus scattering cross section follows an approximate $N_I^2$ scaling.

For a differential neutrino flux $d\phi_{\nu}/dE_\nu$, the differential event rate per unit time per unit detector mass is given by
%%%%
\begin{equation} 
\label{eq:nu_diff_rate}
\dfrac{d R_{\nu}^I}{d E_r} (E_r) = \dfrac{C_I}{m_I} \int^\infty_{E_{\nu}^{\rm min}} \frac{d\phi_{\nu}}{dE_\nu} \dfrac{d \sigma^I}{d E_r} (E_{\nu}, E_r) d E_{\nu}~,
\end{equation}
%%%
where
$E_{\nu}^{\rm min}$ 
is the minimum neutrino energy required to produce a recoil of energy $E_r$, given by
%%%%%%%%
\begin{equation}
E_{\nu}^{\rm min} = \sqrt{\dfrac{m_I E_r}{2}}~.
\end{equation}
%%%%%%%%
Note that the maximum recoil energy due to collision with a neutrino of energy $E_{\nu}$ is
%%%
\begin{equation}
E_r^{\rm max} = \dfrac{2 E_{\nu}^2}{m_I + 2 E_{\nu}}~.
\end{equation}
%%%
In Eq.~\ref{eq:nu_diff_rate}, $C_I$ is the mass fraction of nuclide $I$ in the material. 
When multiple nuclides are present, the differential event rate is obtained by combining their contributions.
The total event rate over the exposure of the experiment is
%%%
\begin{equation} \label{eq:totnurate}
\dfrac{d R_{\nu}}{d E_r} = M \int dT \sum_I \dfrac{d R_{\nu}^I}{d E_r}~,
\end{equation}
%%%
where $M$ is the fiducial target mass of the detector and the integration $\int dT$ runs over the data taking period. 
For neutrino fluxes that are independent of time, these factors can be combined into a single multiplicative term MT that accounts for the total exposure of the experiment. 

Since coherent neutrino-nucleus scattering is mediated by the $Z$ boson, this detection channel is independent of neutrino flavor and the corresponding oscillation effects. 
Oscillation effects do become relevant if neutrino-electron scattering is considered \cite{Gelmini:2018gqa}.

%%%%%%%%%%%%%%%%%
\begin{figure}[tb]
\includegraphics[width=.45\textwidth]{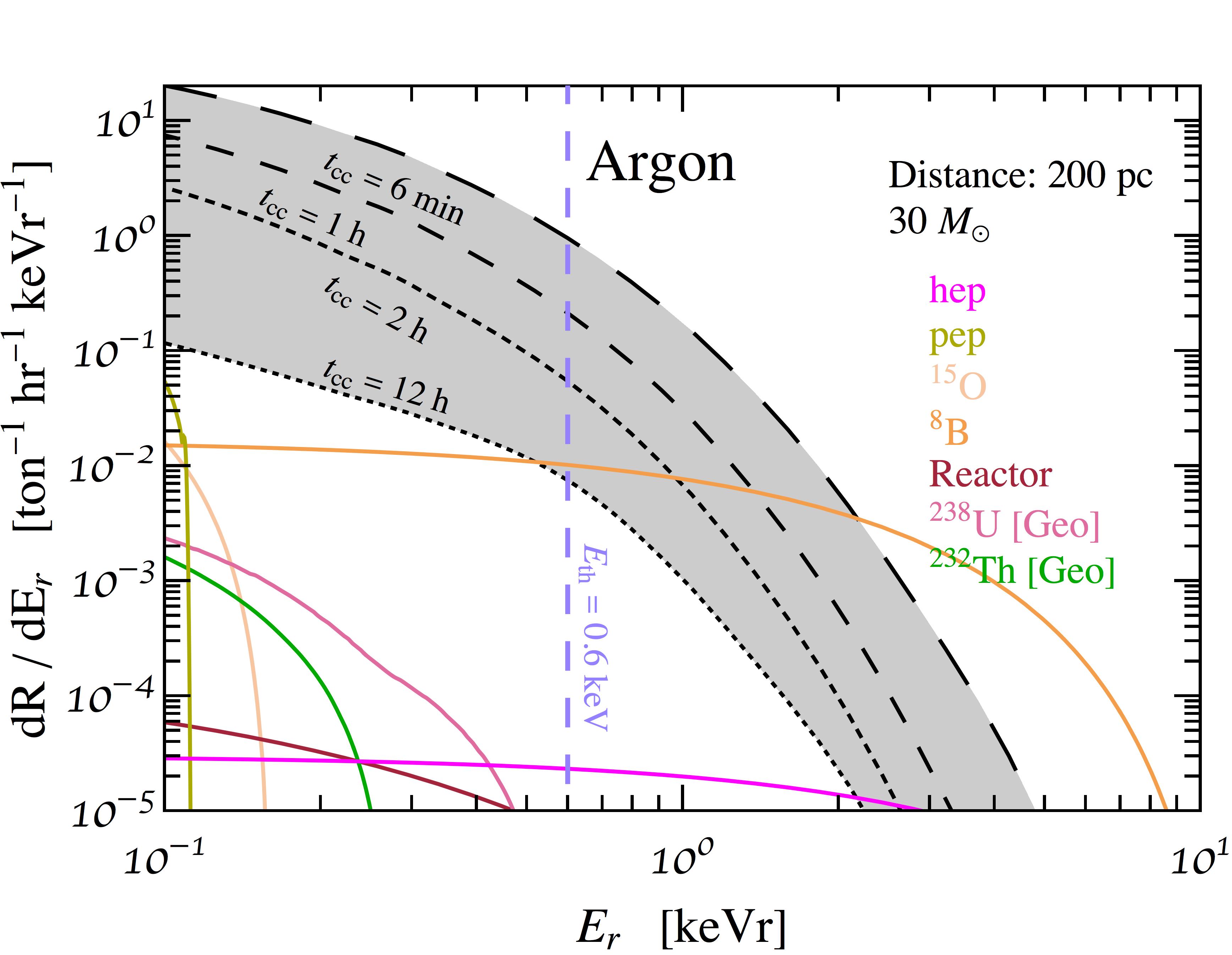}
\caption{\label{fig:nucrec} 
Predicted nuclear recoil spectra for pre-SN neutrino signal in argon for various time intervals until the onset of the core-collapse, assuming a characteristic 30 $M_{\odot}$ star located at a distance of 200 pc. Background contributions from solar, reactor, and geo-neutrinos are also shown for comparison.
}
\end{figure}
%%%%%%%%%%%%%%%%%

We display the expected pre-SN neutrino signal as well as background nuclear recoil spectra for a 300-tonne argon experiment (i.e. ARGO-type detector) in Fig.~\ref{fig:nucrec}.
The pre-SN neutrino signal is evaluated at several time intervals prior to the onset of core-collapse, assuming a characteristic 30 $M_{\odot}$ star located at a distance of 200 pc.
Already around 12 hr prior to core-collapse, the differential recoil rate of the pre-SN neutrino signal near the 0.6 keV detection threshold starts to exceed that of background $^8$B solar neutrinos.

%%%%%%%%%%%%%
\section{Detection Sensitivity} 
\label{sec:sensitivity}
%%%%%%%%%%%%%

A burst of SN-neutrinos is readily detectable with large direct detection experiments for a SN located $\mathcal{O}(10)$ kpc away (i.e. Milky Way distance-scales)~\cite{Chakraborty:2013zua,XMASS:2016cmy,Lang:2016zhv,Kozynets:2018dfo,Khaitan:2018wnf}. 
While argon detectors have not been previously explored as targets for SN neutrinos, the resulting sensitivity can be expected to be similar to xenon-based configurations. 
 In contrast to SN neutrinos, the pre-SN signal is fainter and the detection distance can be expected to be on $\lesssim1$ kpc scales. Nearby red supergiant stars at the end of their lifetime constitute the most likely source for pre-SN neutrino emission. There are 41 red supergiant stars with distance estimates within 1 kpc,
16 within 0.5 kpc, and 5 within 0.2 kpc (see Table 2 of Ref.~\cite{Nakamura:2016kkl}).
The most well studied is the red supergiant Betelgeuse ($\alpha$ Orionis), with mass 17–25 $M_{\odot}$ and distance\footnote{We note that recent parallax measurements obtained from \textit{Gaia} data~\cite{Gaia:2018data} could further decrease the uncertainty in distances to nearby red supergiants (see e.g.~Ref.~\cite{Chatys:2019}).} $197 \pm 45$ pc \cite{Harper:2008}, which we take as our characteristic source. 

%%%%%%%%%%%%%%%%
\begin{figure*}[htb]
\centering
\begin{subfigure}{0.45\textwidth}
  \centering
  \includegraphics[width=\linewidth]{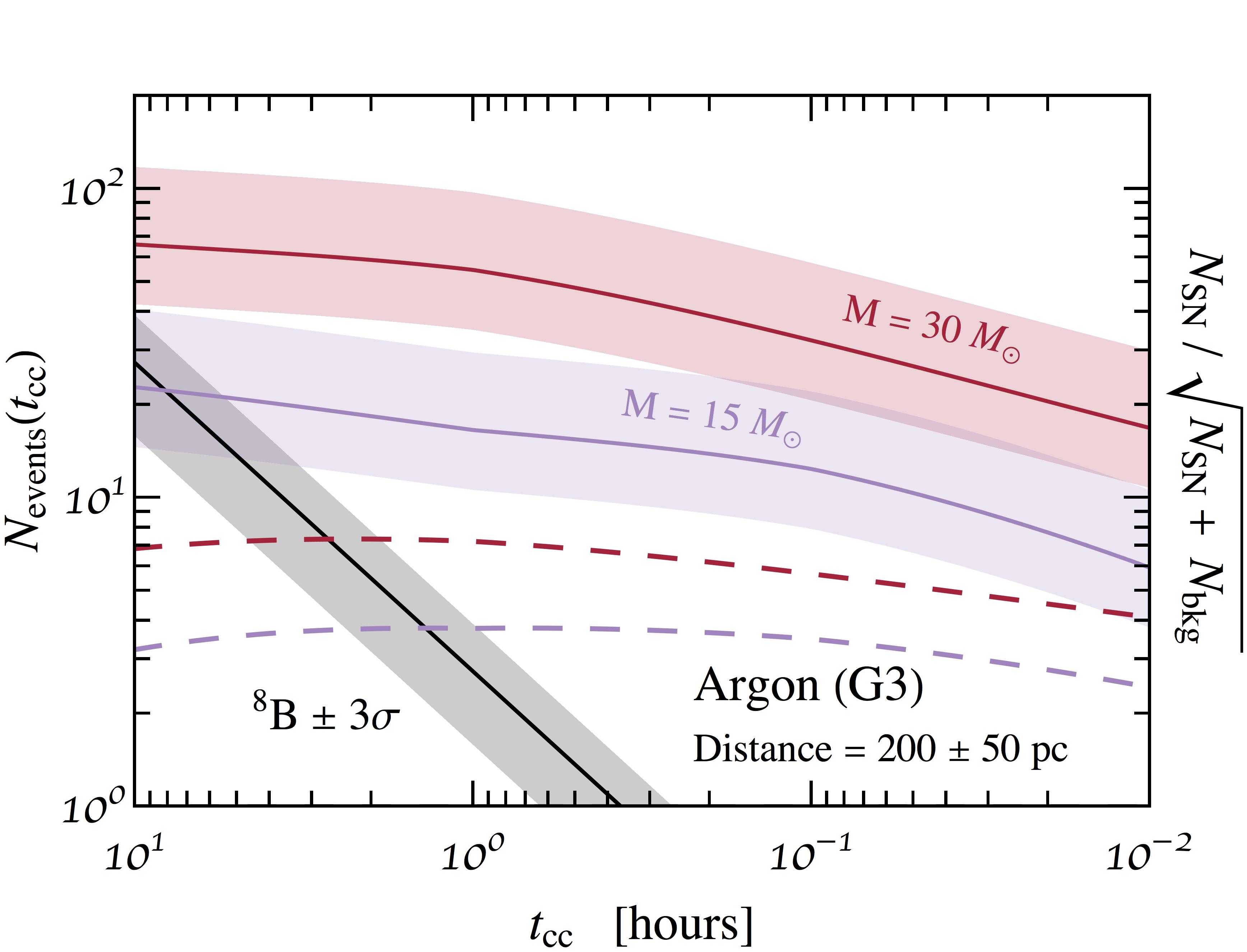}
  \label{fig:sub1}
\end{subfigure}%
\hspace{.4cm}
\begin{subfigure}{0.43\textwidth}
  \centering
  \includegraphics[width=\linewidth]{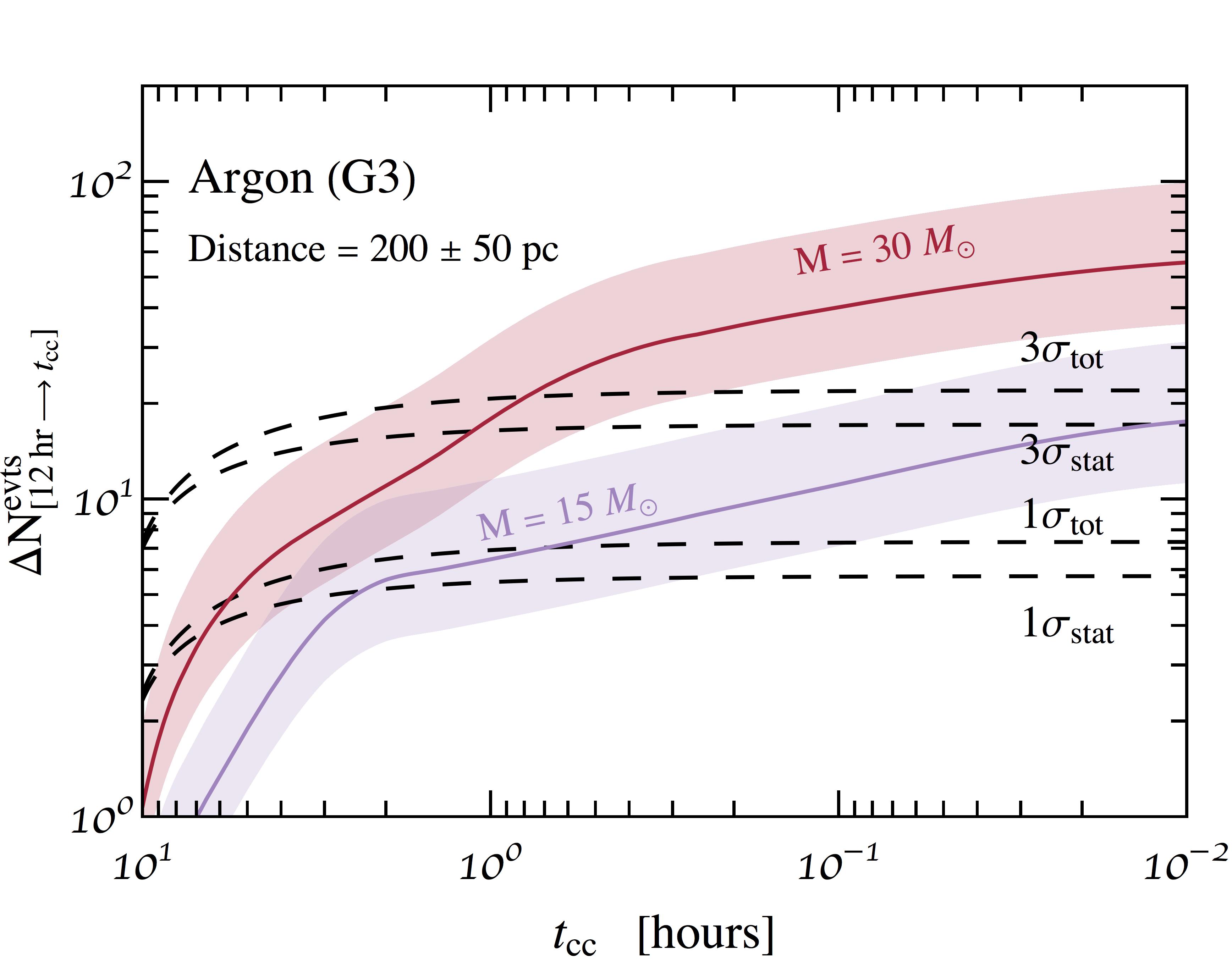}
  \label{fig:sub2}
\end{subfigure}
\caption{Expected number of pre-supernova neutrino events above detection threshold in a generation-3 liquid argon-based dark matter direct detection experiment of target mass 300 tonnes. 
These are shown for stellar masses of $15 M_{\odot}$ and $30 M_{\odot}$ for a star at a distance scale of 200 pc, such as Betelgeuse. 
The error band on pre-supernova neutrino signal is the $\pm1\sigma$ flux uncertainty due to the uncertainty in the star's location.
The errors on the background $^8$B neutrinos come from the systematic uncertainty in the neutrino flux (see Table~\ref{tab:backflux}). [Left] Shown in solid curves are the number of events integrated from the time until core-collapse, $t = -t_{cc}$, to the point of core-collapse, $t = 0$; shown in dashed curves are the ratios of pre-SN events $N_{\rm SN}$ to the square root of $N_{\rm SN} + N_{\rm bkg}$, assuming a SN distance of 200 pc and the $^8$B flux is given by the expected mean value.  
[Right] The number of events integrated from 12 hours before core-collapse, $t$ = -12 hr, to the time until core-collapse, $t = t_{cc}$. Also shown here are 1 $\sigma$ and 3 $\sigma$ statistical and total (statistical and systematic)  fluctuations of the $^8$B neutrino background.}
\label{fig:obsevents}
\end{figure*}
%%%%%%%%%%%%%%%%%%%%%

%%%%%%%%%%%%%%%%
\begin{figure*}[htb]
\centering
\begin{subfigure}{0.45\textwidth}
  \centering
  \includegraphics[width=\linewidth]{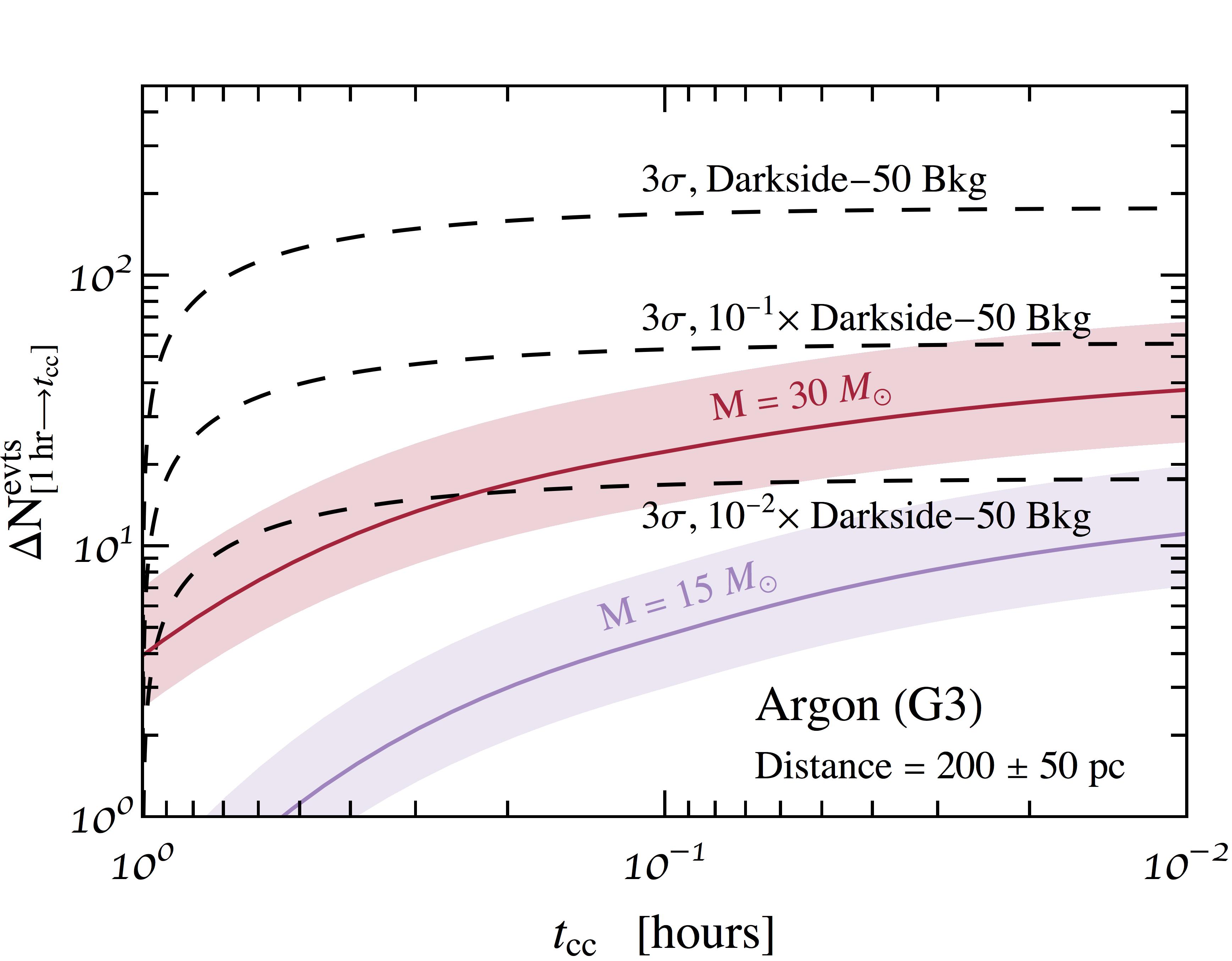}
  \label{fig:sub2}
\end{subfigure}%
\hspace{.4cm}
\begin{subfigure}{0.45\textwidth}
  \centering
  \includegraphics[width=\linewidth]{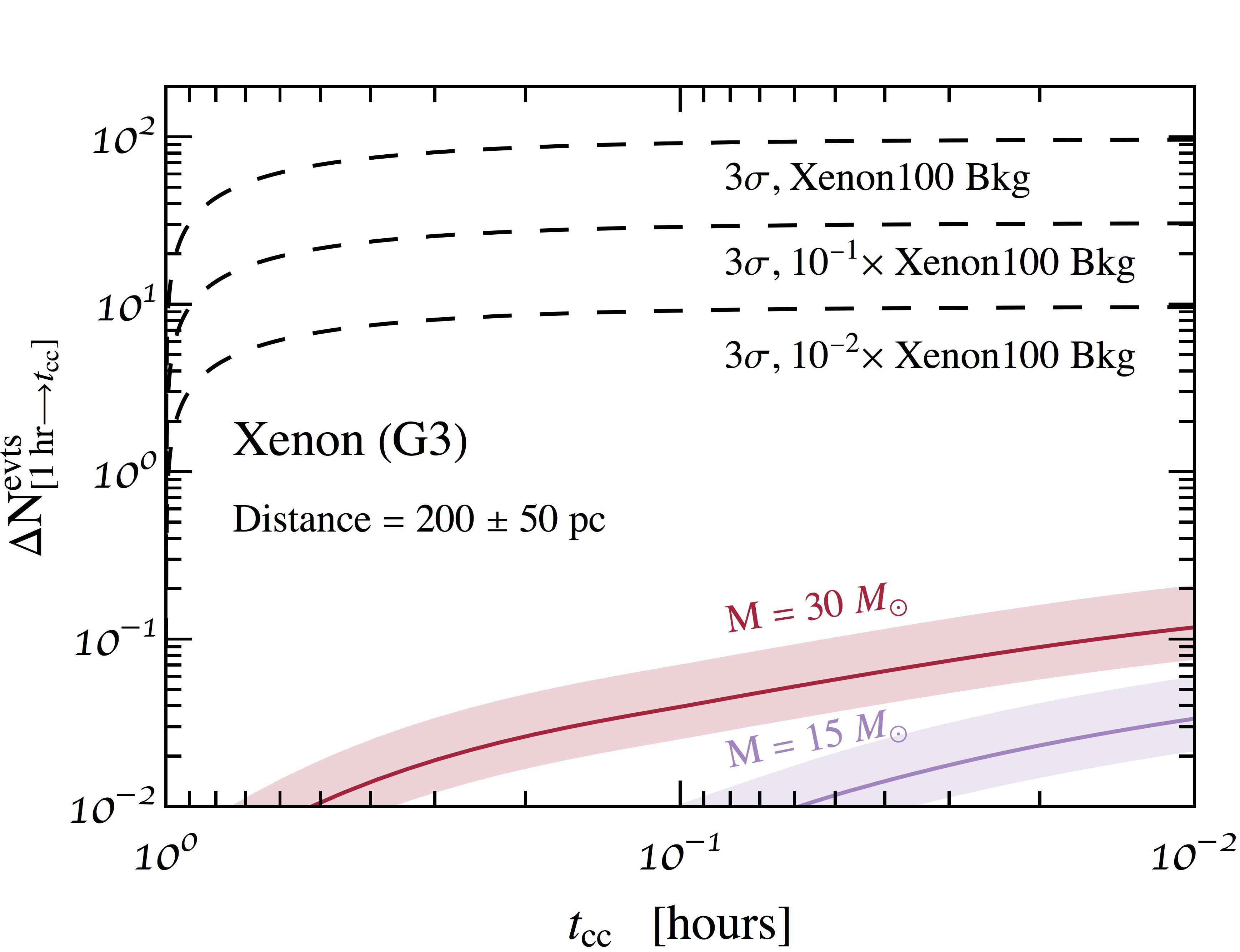}
  \label{fig:sub4}
\end{subfigure}
\caption{Expected number of pre-supernova neutrino events above detection threshold in generation-3 dark matter direct detection experiments with liquid argon target of mass 300 tonnes [left] and liquid xenon target of mass 50 tonnes [right]. 
These are shown for stellar masses of $15 M_{\odot}$ and $30 M_{\odot}$ for a star at a distance of 200 pc.
Also shown are 3 $\sigma$ fluctuations on electronic recoil backgrounds estimated from measurements in~Ref.~\cite{Agnes:2018ves} and~\cite{Aprile:2016wwo} for argon and xenon, respectively (see text for further details).
The error band on pre-supernova neutrino signal is the $\pm1\sigma$ flux uncertainty due to the uncertainty in the star's location.
 }
\label{fig:obseventsERbkg}
\end{figure*}
%%%%%%%%%%%%%%%%%%%%%

%%%%%%%%%%%%%%%%%%%%%
\begin{table*}[tb]
	\begin{center}
		\vspace*{0cm}
		\begin{tabular}[c]{l|ccc|cc|cc} \hline\hline
			Target   & Mass & Threshold & Ref. &  $N_{\rm events}$(12 hr) & $d_{\rm SN}$($N$ = 1) & $N_{\rm events}$(12 hr) & $d_{\rm SN}$($N$ = 1) \\
			& [tonnes] & [keV$_r$] & & $15~M_{\odot}$ & $15~M_{\odot}$ & $30~M_{\odot}$ & $30~M_{\odot}$ \\
			&  & & & [$d = 200$ pc] & [pc] & [$d = 200$ pc] & [pc]   \\ \hline
		argon-G2 & 20 & 0.6 & \cite{Aalseth:2017fik} & 1.53  & 251 & 4.56 & 427 \\ 
     	argon  & 300  & 0.6 & \cite{zuzel2017darkside,Aalseth:2017fik} & 23.6 & 972 & 68.4 & 1654 \\	
     	xenon-G2 & 7 & 0.7 & \cite{Akerib:2015cja,Aprile:2015uzo,Aprile:2016wwo}  & 0.013 & 23.1  & 0.041 & 40.5 \\ 
		  xenon  & 50  & 0.7 & \cite{Aalbers:2016jon,Aprile:2016wwo} & 0.095 & 61.8 &  0.292  & 108 \\
		germanium  & 0.05  & 0.04 & \cite{Agnese:2016cpb} & 0.160 & 78  & 0.356 & 119  \\
		  silicon  & 0.05  & 0.08 & \cite{Agnese:2016cpb} & 0.054 & 46.5 & 0.123 & 70 \\ 
\hline\hline
		\end{tabular}
		\caption{Observational prospects for pre-supernova neutrino signal originating from 15 $M_{\odot}$ and 30 $M_{\odot}$ stars in future large direct dark matter detection experiments with configurations suggested by currently proposed detectors, based on argon (ARGO), xenon (DARWIN), germanium and silicon (both SuperCDMS-SNOLAB), with their respective experimental fiducial mass and threshold.
		The expected number of pre-supernova neutrino events within a 12 hour time window prior to the collapse for a SN 200 pc away from Earth, and  the distance to the source that corresponds to a single mean predicted number of events are displayed. We note that these results are sensitive to assumed level of contributing background.}
		\label{tab:experiments}
		\vspace*{-0.5cm}
	\end{center}
\end{table*}
%%%%%%%%%%%%%%%%%%%%%

We estimate our detection sensitivities in two ways. First, we compute the cumulative number of pre-SN neutrinos events integrated from some time $t = t_{cc}$ prior to core-collapse, all the way to core-collapse, $t = 0$: 
%%%%%%%%
\begin{equation}
N_{\rm events}(t_{cc}) \equiv M \times \int_{-t_{cc}}^0 \, dT \, \int dE_r \, \frac{dR_\nu}{dE_r}~.
\end{equation}
%%%%%%%%
This quantity is displayed on the left-hand panel of Fig.~\ref{fig:obsevents} for two stellar masses, $M = 15 M_{\odot}$ and $M = 30 M_{\odot}$ (shown in solid). Also shown in Fig.~\ref{fig:obsevents} using dashed lines is the ratio of the number of signal (or pre-SN) events $N_{\rm SN}$ to the square root of the total number of events ( $N_{\rm SN}$ +  $N_{\rm bkg}$) as a function of $t_{cc}$, assuming a distance of 200 pc and a $^8$B flux given by the expected mean. It is worth emphasizing that this statistic neglects valuable information on the time dependence of the signal (\ie searches for time-correlated events in a particular window) which could be used to better extract the signal for small $t_{cc}$. The $t_{cc}$ values are chosen to conveniently illustrate the typical average flux behavior. It can be seen that $\mathcal{O}(10)$ pre-SN neutrino events are expected $\sim$1 day prior to the SN explosion, and would exceed the $^8$B neutrino background. This quantity can be extracted from the dataset after the occurrence of the SN, providing a valuable window into the last stages of stellar evolution.

A second useful quantity is the expected number of pre-SN neutrino events obtained in an interval starting $t = $ 12 hours prior to core-collapse and ending at some value $t = t_{cc}$:
%%%%%%%%%
\begin{equation}\label{eq:deltaN}
\Delta N^{\rm evts}_{12 h \rightarrow t_{cc}}(t_{cc}) \equiv M \times \int_{- 12 \, {\rm hr}}^{-t_{cc}} \, dT \, \int dE_r \, \frac{dR_\nu}{dE_r}~.
\end{equation}
%%%%%%%%%%
This is shown in the right-hand panel of Fig.~\ref{fig:obsevents}.
This quantity tracks the number of events accumulated in real time and provides direct visualization of the rising fluence of pre-SN neutrinos. 
In practice, this event-tracking scheme may be implemented within the experiment by registering events over 12-hour windows, with new windows starting every few minutes\footnote{This simple proposal of how event tracking could be implemented may be in some conflict with the currently applied blinding schemes for dark matter searches in which the region of interest is often excluded from the analysis until unblinding.}.
Also shown in the figure are the 1$\sigma$ and 3$\sigma$ statistical and total (statistical plus systematic) fluctuations of the solar $^8$B neutrino background. Note that statistical and systematic uncertainties in the $^8$B flux are comparable for $\Delta N^{\rm evts}_{12 h \rightarrow t_{cc}}(0)$, and thus integrating over a smaller time interval (e.g. instead choosing to analyze, say, $\Delta N^{\rm evts}_{6 h \rightarrow t_{cc}}(0)$) tends to produce subdominant statistical fluctuations. Already at $\mathcal{O}(10)$ hrs prior to collapse, the pre-SN neutrino signal will be visible at 3 $\sigma$ level above the background for a $M = 30 M_{\odot}$ star. For $M = 15 M_{\odot}$ this observation threshold is crossed within an hour of collapse. We note that our projections are conservative and a better future understanding of the $^8$B solar neutrino background will improve the signal sensitivity\footnote{A reduction of $^8$B flux uncertainty by a factor of 2 compared to what we used in our study will allow to detect pre-SN events above 3 $\sigma$ fluctuations of the $^8$B background significantly earlier at $t_{cc} = 1.5$ hours, rather than the displayed $t_{cc} = 0.05$ hours.}.

In ~Fig.~\ref{fig:obseventsERbkg} we display $\Delta N^{\rm evts}_{1 h \rightarrow t_{cc}}(t_{cc})$. defined analogously to $\Delta N^{\rm evts}_{12 h \rightarrow t_{cc}}(t_{cc})$,  
along with 3 $\sigma$ fluctuations associated with the possible electron recoil backgrounds in generation-3 argon and xenon detectors.
We estimate these backgrounds based on measurements by the 
argon-based Darkside-50~\cite{Agnes:2018ves} and 
xenon-based XENON100~\cite{Aprile:2016wwo} experiments. Since electronic recoils can be large at energies above those relevant for pre-SN neutrinos, we compute these background rates using an upper limit on the recoil energy of $1.7$ and $2$ keVnr for xenon and argon, respectively. Over the energy range of interest these rates are approximately independent of recoil energy, being $\sim 0.2$ (argon) and $\sim 0.5$ (xenon) events/keVnr/kg/day. We consider the more pessimistic case where these backgrounds are not reduced when generation-3 experiments come online, as well as the more optimistic one where they are reduced by a factor of 10 or 100 at such a time.

In Table~\ref{tab:experiments} we list the expected number of observed pre-SN neutrino events within a 12 hour period prior to the collapse, and also the distance from the source at which a single event is observable, for the proposed large scale direct detection configurations we consider.
Due to potentially larger achievable target mass and lower threshold, an argon-based experiment could be more sensitive to pre-SN neutrinos than an experiment based on xenon. Despite potentially lower thresholds for germanium- and silicon-based configurations, we expect that an argon-based experiment could constitute a significantly more favorable pre-SN neutrino detector due to the much greater target mass. 

\begin{figure*}[htb]
\centering
\begin{subfigure}{0.5\textwidth}
  \centering
  \includegraphics[width=\linewidth]{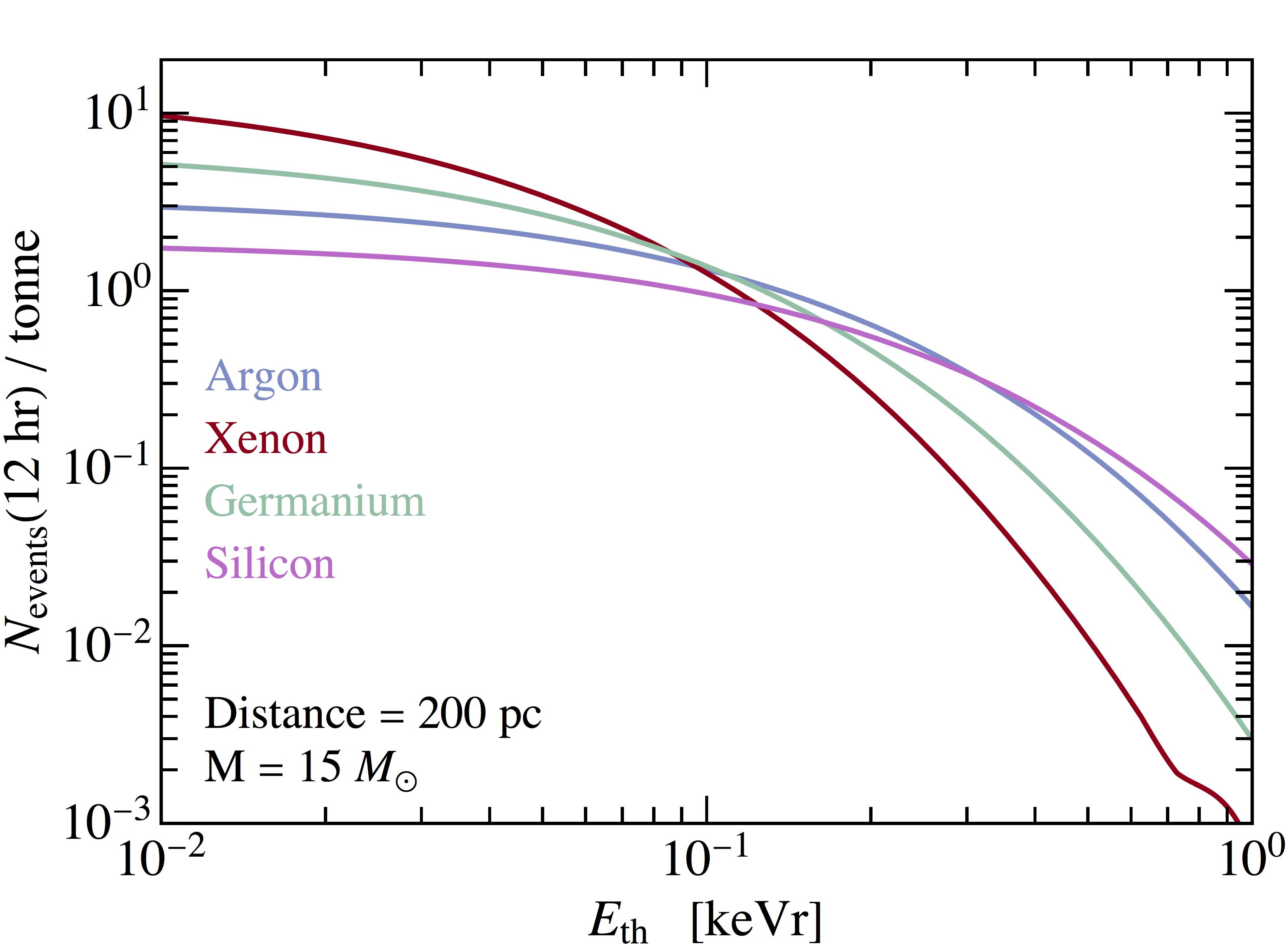}
  \label{fig:sub1}
\end{subfigure}%
\begin{subfigure}{0.5\textwidth}
  \centering
  \includegraphics[width=\linewidth]{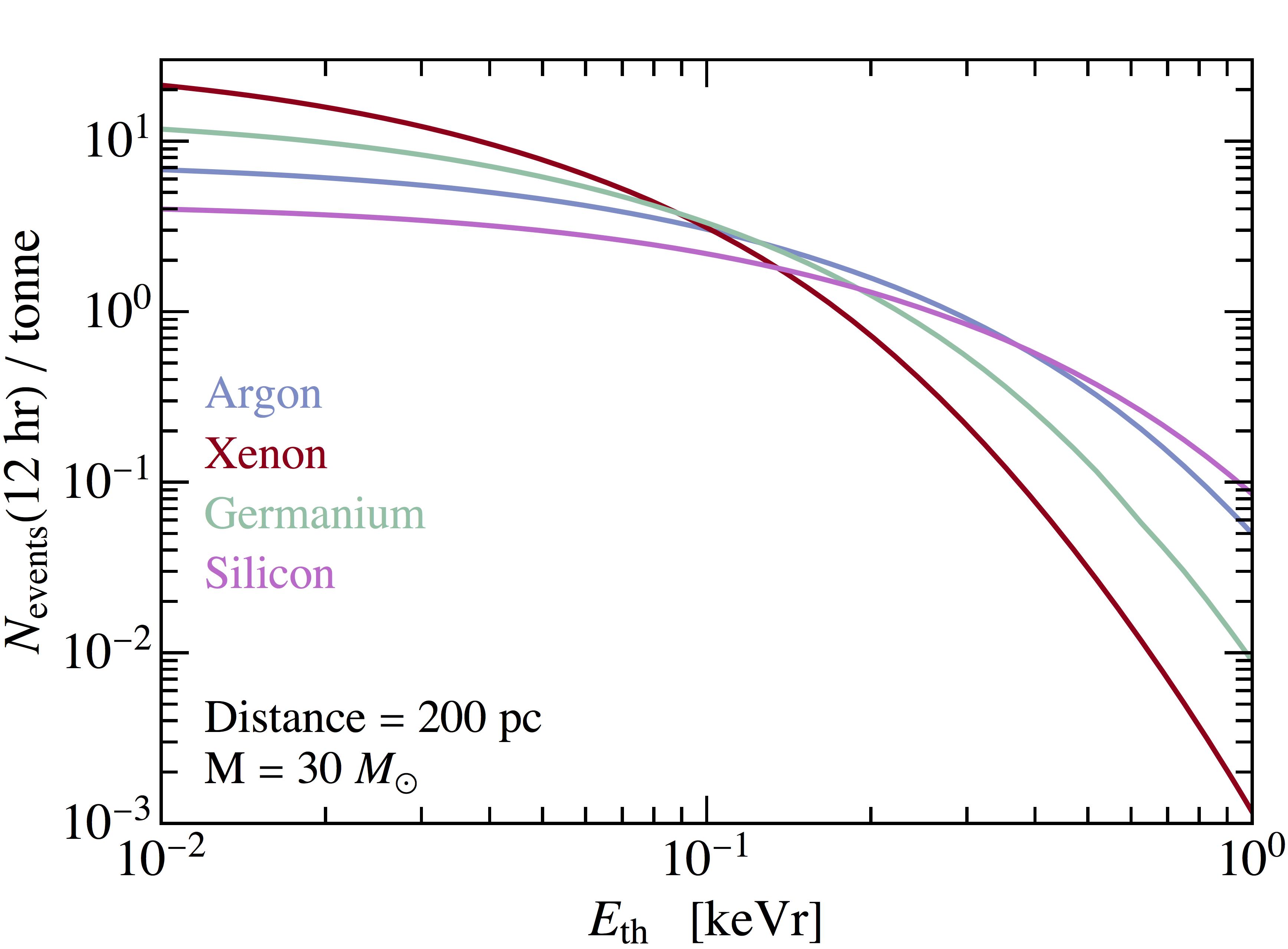}
  \label{fig:sub2}
\end{subfigure}
\caption{Expected number of events per unit target mass as a function of energy threshold, for detectors with argon, xenon, germanium, and silicon, assuming a 15 $M_\odot$ (left) and 30 $M_\odot$ (right) star at a distance of $200$ pc.}
\label{fig:thresh}
\end{figure*}

We note that by significantly improving the detection threshold
for a given target material, similar event rates can be achieved with smaller exposures. In \Fig{fig:thresh}, we illustrate how the sensitivity depends on the adopted threshold (taking the maximum recoil energy to be 10 keV) for each of our experimental configurations, for a 15$M_\odot$ (left) and 30$M_\odot$ (right) star at 200 pc. We display here the number of events per unit target mass (in units of ${\rm tonne}^{-1}$) to allow for a straightforward comparison between experiments.

%%%%%%%%%%%%%%%%%%
\section{Summary} 
\label{sec:summary}
%%%%%%%%%%%%%%%%%

A future supernova is a highly anticipated event, and experiments that can detect the corresponding pre-supernova neutrinos can both serve as efficient supernova alarm triggers and provide novel insights into the final stages of the lifetime of massive stars. Future large-scale dark matter direct detection experiments can also act as effective neutrino telescopes. In this work we have explored their sensitivity to pre-supernova neutrino signal detection.

Unlike conventional neutrino experiments that primarily employ inverse beta decay for SN-related neutrinos, direct detection experiments will detect these neutrinos primarily through the channel of coherent nuclear scattering, which has many advantages. First, the event rate is independent of neutrino mass hierarchy and oscillations, and in particular there is no penalty to pay in the case of inverted hierarchy~\cite{Asakura:2015bga}. Second, since all neutrino flavors participate in coherent nuclear scattering, the pre-supernova neutrino flux that could be detected in direct detection experiments is potentially larger compared to typical neutrino experiments. We do note, however, that the direct detection experiments we consider are insensitive to the directionality of the pre-supernova neutrinos\footnote{However, directionality may be challenging for neutrino experiments as well, due to IBD detection techniques as well as the low energies of pre-supernova neutrinos.}.

In summary, we have shown that large-scale dark matter direct detection experiments, especially those based on argon and xenon, can constitute promising pre-supernova neutrino detectors. Large scale direct detection experiments thus complement dedicated neutrino telescopes as pre-supernova detectors. Our work also highlights the necessity for further efforts to reduce experimental backgrounds at lower thresholds in direct detection experiments that will allow to open sensitivity for new programs of research.

\section*{Acknowledgements}

We would like to thank 
Joe Bramante,
Michela Lai, 
Rafael Lang,
Ingrida Semenec, 
Michael Smy,
Hank Sobel,
Mark Vagins,
and
Tien-Tien Yu
for helpful discussions.
The work of NR is supported by the Natural Sciences and Engineering Research Council of Canada; TRIUMF receives federal funding via a contribution agreement
with the National Research Council Canada. 
The work for VT was supported by the U.S. Department of Energy (DOE) Grant No. DE-SC0009937.
SJW acknowledges support from Spanish MINECO national grants FPA2014-57816-P and FPA2017-85985-P, and from the European projects H2020-MSCAITN-2015//674896-ELUSIVES and H2020-MSCA-RISE2015.

\bibliography{presnlib}
\end{document}